


\input lanlmac
\input amssym
\input epsf

\newcount\figno
\figno=0
\def\fig#1#2#3{
\par\begingroup\parindent=0pt\leftskip=1cm\rightskip=1cm\parindent=0pt
\baselineskip=13pt
\global\advance\figno by 1
\midinsert
\epsfxsize=#3
\centerline{\epsfbox{#2}}
\vskip 12pt
{\bf Fig. \the\figno:~~} #1 \par
\endinsert\endgroup\par
}
\def\figlabel#1{\xdef#1{\the\figno}}


\def\th{\theta}

\def\ep{\epsilon}

\def\Z{{\bf Z}}

\def\hf{{1\over 2}}
\def\qu{{1\over 4}}

\def\o{\over}

\def\til#1{\widetilde{#1}}
\def\si{\sigma}
\def\Si{\Sigma}
\def\b#1{\overline{#1}}
\def\del{\partial}

\def\lap{\Delta}
\def\bra{\langle}
\def\ket{\rangle}
\def\lf{\left}
\def\ri{\right}
\def\riya{\rightarrow}

\def\Ga{\Gamma}

\def\om{\omega}

\def\rt#1{\sqrt{#1}}

\def\sitarel#1#2{\mathrel{\mathop{\kern0pt #1}\limits_{#2}}}
\def\uerel#1#2{{\buildrel #1 \over #2}}

\def\cob{\delta}

\lref\KarczmarekXM{
  J.~L.~Karczmarek, H.~Liu, J.~Maldacena and A.~Strominger,
  ``UV finite brane decay,''
  JHEP {\bf 0311}, 042 (2003)
  [arXiv:hep-th/0306132].
} \lref\LukyanovNJ{
  S.~L.~Lukyanov, E.~S.~Vitchev and A.~B.~Zamolodchikov,
  ``Integrable model of boundary interaction: The paperclip,''
  Nucl.\ Phys.\ B {\bf 683}, 423 (2004)
  [arXiv:hep-th/0312168].
} \lref\LukyanovBF{
  S.~L.~Lukyanov and A.~B.~Zamolodchikov,
  ``Dual form of the paperclip model,''
  arXiv:hep-th/0510145.
} \lref\StromingerPC{
  A.~Strominger,
  ``Open string creation by S-branes,''
  arXiv:hep-th/0209090.
} \lref\KutasovRR{
  D.~Kutasov,
  ``Accelerating branes and the string / black hole transition,''
  arXiv:hep-th/0509170.
} \lref\RibaultSS{
  S.~Ribault and V.~Schomerus,
  ``Branes in the 2-D black hole,''
  JHEP {\bf 0402}, 019 (2004)
  [arXiv:hep-th/0310024].
} \lref\NakayamaPK{
  Y.~Nakayama, S.~J.~Rey and Y.~Sugawara,
  ``D-brane propagation in two-dimensional black hole geometries,''
  JHEP {\bf 0509}, 020 (2005)
  [arXiv:hep-th/0507040].
} \lref\MaloneyCK{
  A.~Maloney, A.~Strominger and X.~Yin,
  ``S-brane thermodynamics,''
  JHEP {\bf 0310}, 048 (2003)
  [arXiv:hep-th/0302146].
} \lref\DijkgraafBA{
  R.~Dijkgraaf, H.~L.~Verlinde and E.~P.~Verlinde,
  ``String propagation in a black hole geometry,''
  Nucl.\ Phys.\ B {\bf 371}, 269 (1992).
} \lref\OkudaFB{
  T.~Okuda and T.~Takayanagi,
  ``Ghost D-branes,''
  arXiv:hep-th/0601024.
} \lref\DouglasUP{
  M.~R.~Douglas, I.~R.~Klebanov, D.~Kutasov, J.~Maldacena, E.~Martinec and N.~Seiberg,
  ``A new hat for the c = 1 matrix model,''
  arXiv:hep-th/0307195.
} \lref\GaiottoYF{
  D.~Gaiotto, N.~Itzhaki and L.~Rastelli,
  ``On the BCFT description of holes in the c = 1 matrix model,''
  Phys.\ Lett.\ B {\bf 575}, 111 (2003)
  [arXiv:hep-th/0307221].
} \lref\SusskindIF{
  L.~Susskind, L.~Thorlacius and J.~Uglum,
  ``The Stretched horizon and black hole complementarity,''
  Phys.\ Rev.\ D {\bf 48}, 3743 (1993)
  [arXiv:hep-th/9306069].
}
\lref\FidkowskiNF{
  L.~Fidkowski, V.~Hubeny, M.~Kleban and S.~Shenker,
  JHEP {\bf 0402}, 014 (2004)
  [arXiv:hep-th/0306170].
} 
\lref\MaldacenaKR{
  J.~M.~Maldacena,
  JHEP {\bf 0304}, 021 (2003)
  [arXiv:hep-th/0106112].
}
\lref\IsraelUR{
  W.~Israel,
  Phys.\ Lett.\ A {\bf 57}, 107 (1976).
}\lref\StromingerPC{
  A.~Strominger,
  ``Open string creation by S-branes,''
  arXiv:hep-th/0209090.
} \lref\NakayamaYX{
  Y.~Nakayama, Y.~Sugawara and H.~Takayanagi,
  ``Boundary states for the rolling D-branes in NS5 background,''
  JHEP {\bf 0407}, 020 (2004)
  [arXiv:hep-th/0406173].
} \lref\GinspargIS{
  P.~H.~Ginsparg and G.~W.~Moore,
  ``Lectures on 2-D gravity and 2-D string theory,''
  arXiv:hep-th/9304011.
} \lref\GutperleXF{
  M.~Gutperle and A.~Strominger,
  ``Timelike boundary Liouville theory,''
  Phys.\ Rev.\ D {\bf 67}, 126002 (2003)
  [arXiv:hep-th/0301038].
} \lref\MartinecQT{
  E.~Martinec and K.~Okuyama,
  ``Scattered results in 2D string theory,''
  JHEP {\bf 0410}, 065 (2004)
  [arXiv:hep-th/0407136].
} \lref\KarczmarekBW{
  J.~L.~Karczmarek, J.~Maldacena and A.~Strominger,
  ``Black hole non-formation in the matrix model,''
  arXiv:hep-th/0411174.
} \lref\KazakovPM{
  V.~Kazakov, I.~K.~Kostov and D.~Kutasov,
  ``A matrix model for the two-dimensional black hole,''
  Nucl.\ Phys.\ B {\bf 622}, 141 (2002)
  [arXiv:hep-th/0101011].
} \lref\KutasovFG{
  D.~Kutasov, K.~Okuyama, J.~w.~Park, N.~Seiberg and D.~Shih,
  ``Annulus amplitudes and ZZ branes in minimal string theory,''
  JHEP {\bf 0408}, 026 (2004)
  [arXiv:hep-th/0406030].
} \lref\MaldacenaHI{
  J.~Maldacena,
  ``Long strings in two dimensional string theory and non-singlets in the
  matrix model,''
  JHEP {\bf 0509}, 078 (2005)
  [arXiv:hep-th/0503112].
} \lref\GaiottoGD{
  D.~Gaiotto,
  ``Long strings condensation and FZZT branes,''
  arXiv:hep-th/0503215.
} \lref\MooreZV{
  G.~W.~Moore, M.~R.~Plesser and S.~Ramgoolam,
  ``Exact S matrix for 2-D string theory,''
  Nucl.\ Phys.\ B {\bf 377}, 143 (1992)
  [arXiv:hep-th/9111035].
} \lref\KutasovDJ{
  D.~Kutasov,
  ``D-brane dynamics near NS5-branes,''
  arXiv:hep-th/0405058.
} \lref\SahakyanCQ{
  D.~A.~Sahakyan,
  ``Comments on D-brane dynamics near NS5-branes,''
  JHEP {\bf 0410}, 008 (2004)
  [arXiv:hep-th/0408070].
} \lref\YogendranDM{
  K.~P.~Yogendran,
  ``D-branes in 2D Lorentzian black hole,''
  JHEP {\bf 0501}, 036 (2005)
  [arXiv:hep-th/0408114].
}
\lref\SenYV{
  A.~Sen,
  ``Symmetries, conserved charges and (black) holes in two dimensional  string
  theory,''
  JHEP {\bf 0412}, 053 (2004)
  [arXiv:hep-th/0408064].
}
\lref\YogendranDM{
  K.~P.~Yogendran,
  ``D-branes in 2D Lorentzian black hole,''
  JHEP {\bf 0501}, 036 (2005)
  [arXiv:hep-th/0408114].
}
\lref\WittenYR{
  E.~Witten,
  ``On string theory and black holes,''
  Phys.\ Rev.\ D {\bf 44}, 314 (1991).
}
\Title{
                                             \vbox{
                                             \hbox{hep-th/0602060}}}
{\vbox{ \centerline{Hairpin Branes and D-Branes Behind the
Horizon} }}

\vskip .2in

\centerline{Kazumi Okuyama$^1$ and Moshe Rozali$^{1,2}$}

\vskip .2in

\centerline{$^1$ Department of Physics and Astronomy}\centerline{
University of British Columbia} \centerline{Vancouver, BC, V6T
1Z1, Canada}  \centerline{$^2$ Perimeter Institute for Theoretical
Physics}\centerline{ Waterloo, Ontario, N2L 2Y5, Canada}
 \centerline{{\tt kazumi@phas.ubc.ca,
rozali@phas.ubc.ca}}

\vskip 2cm \noindent

We study Lorentzian D-particles in linear dilaton and the two
dimensional black hole backgrounds.  The D-particle trajectory
follows an accelerated trajectory which is  smeared by stringy
corrections. For the black hole background we find that the portion
of the trajectory behind the horizon appears  to an asymptotic
observer as ghost D-particle. This suggests a way of constructing a
matrix model for the Lorentzian black hole background.\Date{February
2006}

\vfill
\vfill

\newsec{Introduction}

Time dependence, and more generally Lorentzian physics is an
interesting and important subject in string theory. Generally
calculations in string perturbation theory are performed in
Euclidean space, resulting in ambiguities once analytic
continuation to Lorentzian signature space is performed. These
ambiguities are physically important, encoding  for example  such
issues as vacuum choice and Hawking radiation from black holes. It
is interesting therefore to investigate this issue with an eye on
the differences between string theory and ordinary quantum field
theory in curved spacetime.

 In this context, the hairpin branes, the topic of this paper, are
interesting objects since they are time dependent and have an
exact (tree level) worldsheet definition. The hairpin branes are
defined in the linear dilaton backgrounds and their exact boundary
states are known explicitly \LukyanovNJ. At the classical level,
the hairpin brane is described by the boundary condition that the
open string is attached to a fixed locus, which describes an
hairpin shape. This picture is modified by $\alpha'$ corrections
that can be treated exactly. As noted in \KutasovRR, the classical
hairpin shape is smeared due to the presence of open string
tachyon condensate near the tip of Euclidean hairpin. In a recent
paper \LukyanovBF, this picture is made more precise by showing
that the boundary CFT of hairpin brane is T-dual to a system with
Liouville-like boundary potential. This can be thought of as a
boundary analogue of the cigar/sine-Liouville duality.
Interestingly, as we discuss in the text, even after including the
worldsheet effects the exact boundary state has an expression as a
superposition of classical trajectories with some smearing factor.
This interplay between the classical trajectory and the exact
boundary state is one of the main observations in this paper.

Upon Wick rotation to the Lorentzian signature, the hairpin brane
becomes an accelerating brane, which moves towards the strong
coupling end of the linear dilaton direction
\refs{\KutasovDJ,\NakayamaYX,\SahakyanCQ, \KutasovRR}. Since it is
a simple and clean time-dependent system, we can address many
interesting questions related to the time-dependence in string
theory. In particular, we consider the closed string field sourced
by the brane and the closed string emission from the Lorentzian
hairpin brane (which we find to be finite). We find that the
classical Rindler horizon that exists for any accelerated
trajectory  is modified due to the stringy smearing of the D-brane
trajectory.

There also exists an analogue of hairpin brane in two dimensional
black hole background. As in the hairpin brane case, the exact
boundary state on the Euclidean black hole is known explicitly
\RibaultSS. Again, the exact disk one-point function has a simple
relation to the classical trajectory of brane \NakayamaPK.
However, in contrast to the hairpin brane case, the Wick rotation
to the Lorentzian signature is a non-trivial problem due to the
presence of horizons in the Lorentzian black hole. The Euclidean
section of black hole geometry only covers a causal patch outside
the horizon in the Lorentzian geometry, thus the extension of the
boundary state beyond the horizon is not unique. In this paper, we
propose a prescription how to extend the boundary state to the
entire Lorentzian black hole geometry, which amounts to imposing
"transparency" when the D-particle crosses the horizon. Our method
relies  on the relation between the classical trajectory and the
exact boundary state. Surprisingly, we find that the D-brane
behind the horizon is a ``ghost D-brane'' introduced recently in
\OkudaFB.

This paper is organized as follows. In section 2, we consider
hairpin branes in flat  Euclidean space and their Wick rotation to
the Lorentzian space. We find an interesting interplay between the
classical trajectory and the exact boundary state. Then we study
the closed string emission from the Lorentzian hairpin brane and
find that the emission rate is UV finite as in case of the rolling
tachyon in the linear dilaton background.

In section 3, we consider D0-branes in 2d black hole background.
Using the relation of classical trajectory and exact boundary
state, we extend the definition of boundary state from one causal
patch to the entire black hole geometry. In particular, we find
that the D0-brane behind the horizon is a ghost D-brane. We also
consider the closed string emission from D0-branes and discuss its
dependence on the choice of closed string vacuum.

Section 4 is devoted to the discussion of possible significance of
the ghost D-branes behind the horizon. Based on our result we
speculate a possible matrix model dual of 2d Lorentzian black hole
involving a supergroup.

\newsec{Hairpin Brane}

In this section we discuss the hairpin brane in flat space with a
linear dilaton background. We start by discussion of the features of
the Euclidean brane, followed by a discussion of Lorentzian features
such as the closed string field and the (open and closed) string
radiation emitted from the brane.

\subsec{Euclidean Hairpin}

Let us first consider the Euclidean
version of hairpin brane. This is a boundary state in a system of
two bosons $X,Y$ (and possibly some additional compact CFT as needed
to provide critical string background). The system has the linear
dilaton background in the $X$ direction \eqn\dilaton{ \Phi=-{Q\o2}X.
} The energy momentum tensor is \eqn\TXY{ T=-\hf(\del
X)^2-{Q\o2}\del^2X-\hf(\del Y)^2 } and the central charge is
$c=2+3Q^2$. The conformal weight of the tachyon vertex operator
$e^{-{Q\o2}X+i{\bf p}\cdot {\bf X}}$ is $\lap={Q^2\o8}+\hf{\bf
p}^2$. Note also that the tachyons with positive and negative
momenta $p_x$ are unrelated, unlike the situation in Liouville
theory where they are related by the reflection coefficient.

The hairpin brane is defined by the boundary condition
\eqn\hairpinshape{ e^{-{Q\o2}X}= C \cos{Y\o2a} }
 where  $C$ is a constant characterizing the trajectory, and $a$ is given by \eqn\ainQ{ a=\rt{1+{1\o Q^2}} } We see that
in the weak coupling region $X\sim+\infty$, $Y$ approaches $Y=\pm
\pi a$, hence the name "hairpin" brane. The coordinate $Y$ is
non-compact, but the trajectory extends only in the interval $|Y|
\leq \pi a$, thus there is no periodicity in the Euclidean time
direction.

The boundary state for the hairpin brane is \eqn\Bstate{ \bra
B_{\subset}({\bf a})| =\int d^2p\,\Psi_{\subset}({\bf p},{\bf
a})\bra\bra{\bf p}| } where  $\bra\bra{\bf p}|$ is the Ishibashi
state built on the primary with momentum ${\bf p}$. The modulus $\bf
a$ of the boundary state characterizes the hairpin's tip location in
spacetime (and should not be confused with the constant $a$). We
will for the most part choose coordinates such that $\bf a =0$,
which chooses a specific value for the constant $C$ in  the boundary
condition \hairpinshape.

The coefficients $\Psi_{\subset}({\bf p},{\bf a})$ summarize the
one point functions for the closed strings sourced by the brane.
They are given by \eqn\wavefn{ \Psi_{\subset}({\bf p},{\bf
a})=e^{i{\bf p}\cdot{\bf a}} {{1\o Q}\Ga(-2i{p_x\o
Q})\Ga(1-ip_xQ)\o \Ga(\hf-i{p_x\o Q}-ap_y)\Ga(\hf-i{p_x\o
Q}+ap_y)} } where we omit an irrelevant overall constant for
simplicity. In the semiclassical limit $Q \rightarrow 0$ the
Fourier transform of the wavefunction $\Psi_{\subset}$ localizes
on the trajectory \hairpinshape. The factor $\Ga(-2i{p_x\o Q})$ is
interpreted as the effect of the bending of the brane
\hairpinshape\ represented by the screening charge $e^{-{Q\o2}X}$
\KutasovRR. On the other hand, the factor $\Ga(1-ip_xQ)$
represents the effect of the open string tachyon. Indeed, it is
shown that the locations of the poles of $\Ga(1-ip_xQ)$ are
reproduced by the insertion of the screening operator \LukyanovBF
\eqn\Sbdry{ S_B=\int_{\del\Si}d\tau \,e^{-{X\o
Q}}(\si_+e^{ia\til{Y}}+\si_-e^{-ia\til{Y}}) }
 where $\si_{\pm}$ are Pauli matrices.

 An interesting observable is the annulus amplitude which can be
  calculated in the closed string channel    \eqn\annEuc{\eqalign{ Z({\bf a}|\,{\bf a}')&=\bra
B_{\subset}({\bf a})| \,q_c^{\hf(L_0+\tilde{L}_0)-{c\o24}}
|B_{\subset}({\bf a}')\ket \cr &={1\o4}\int d^2p\,e^{i{\bf
p}\cdot({\bf a}-{\bf a}')}{\cos(2\pi ap_y) +\cosh({2\pi p_x\o Q})\o
\sinh(\pi p_xQ)\sinh({2\pi p_x\o Q})} {1\o \eta(q_c)^2}q_c^{\hf{\bf
p}^2} } }Here $q_c=e^{-2\pi t_c}$  is the annulus modular parameter
in the closed string channel, and the factor ${1\o
\eta(q_c)^2}q_c^{\hf{\bf p}^2}$ is the character of the Ishibashi
state with momentum ${\bf p}$. Note that the characters in the
bosonic case are known explicitly, a fact we will find useful below.

One can perform a modular transformation to the open string channel,
with $ q_o=e^{-2\pi/t_c}$ being the open string modular parameter.
We make use of the formula \eqn\formeta{
\int{d^Dx\o(2\pi)^D}e^{i{\bf p}\cdot{\bf x}}
{1\o\eta(q_o)^D}q_o^{{1\o8\pi^2}{\bf x}^2}
={1\o\eta(q_c)^D}q_c^{\hf{\bf p}^2}. } giving the open string
expression\eqn\annEuc{ Z({\bf a}|\,{\bf
a}')={1\o4}\int{d^2x\,d^2p\o(2\pi)^2}e^{i{\bf p}\cdot {\bf
x}}{\cos(2\pi ap_y) +\cosh({2\pi p_x\o Q})\o \sinh(\pi
p_xQ)\sinh({2\pi p_x\o Q})} {1\o\eta(q_o)^2}q_o^{{1\o 8\pi^2}({\bf
x}-{\bf a}+{\bf a}')^2} }

It is natural to classify the different terms in the annulus
amplitude as coming from open strings stretched between either the
same side or different sides of the hairpin. Indeed, the $p_y$
integral leads to terms proportional to the $\cob$-functions
\eqn\pyint{ \cob(y \pm 2\pi a)~~~~  or ~~~~\quad \cob(y). }

The first pair of  $\cob$-functions correspond to the open string
stretched between the opposite sides of the D-brane $Y=\pm \pi a$,
while $\cob(y)$ comes from the open strings on the same side. This
suggests a definition of "half-hairpin" branes corresponding to one
of the branches of the hairpin only. To that end  we can define the
wave functions \eqn\wavehalf{ \Psi_{\Longleftarrow}({\bf p},{\bf
a})=e^{i{\bf p}\cdot{\bf a}} {{1\o \sqrt{2} \,Q}\,\Ga(-2i{p_x\o
Q})\,\Ga(1-ip_xQ) \, \exp({{\pi p_x \o Q}+ \pi i a p_y } )}} for the
first "incoming" branch and  \eqn\wavehalf{
\Psi_{\Longrightarrow}({\bf p},{\bf a})=e^{i{\bf p}\cdot{\bf a}}
{{1\o \sqrt{2} \,Q}\,\Ga(-2i{p_x\o Q})\,\Ga(1-ip_xQ) \, \exp({{\pi
p_x \o Q}- \pi i a p_y } )}} for the second "outgoing" branch. With
these definitions the terms proportional to $\cob(y)$ come from
\eqn\annhalfone{ \bra B_{\Longrightarrow}({\bf a})|
\,q_c^{\hf(L_0+\tilde{L}_0)-{c\o24}} |B_{\Longrightarrow}({\bf
a}')\ket~~~~or~~~~ \bra B_{\Longleftarrow}({\bf a})|
\,q_c^{\hf(L_0+\tilde{L}_0)-{c\o24}} |B_{\Longleftarrow}({\bf
a}')\ket} whereas the terms proportional to $\cob(y \pm 2\pi a)$
come from \eqn\annhalftwo{ \bra B_{\Longleftarrow}({\bf a})|
\,q_c^{\hf(L_0+\tilde{L}_0)-{c\o24}} |B_{\Longrightarrow}({\bf
a}')\ket~~~~or~~~~ \bra B_{\Longrightarrow}({\bf a})|
\,q_c^{\hf(L_0+\tilde{L}_0)-{c\o24}} |B_{\Longleftarrow}({\bf
a}')\ket}
 Note that the half branes $|B_{\Longleftarrow}\ket,|B_{\Longrightarrow}\ket$
are localized at $Y=\pm\pi a$.
When Wick rotated to the Lorentzian signature, the half branes are
 analogous to a purely decaying D-branes and its time reversal
(half S-branes \GutperleXF). We note that these boundary states
satisfy the Cardy conditions separately.

 These boundary states can be also constructed using the T-dual picture. Indeed,
the Euclidean hairpin brane is T-dual to the system with boundary
potential \Sbdry\ \LukyanovBF.
Therefore the half
branes we define correspond, as for the rolling tachyon case, to
boundary interactions including only one of the exponentials. We
limit ourselves in what follows to discussion of the time-symmetric
brane only .

Returning to the partition function \annEuc\  in the open string
variables, the $p_x$ integral can be performed as in eq.(4.11) of
\KarczmarekXM. When ${\bf a}={\bf a}'=0$, the amplitude turns out to be
a sum of two terms
\eqn\ZIzeroone{ Z=\hf(I_0+I_1), }
\eqn\Izeroone{
I_0=\int_{-\infty}^\infty ds\rho_0(s)
 {q^{\hf s^2}_o\o\eta(q_o)^2},\quad
I_1=\int_{-\infty}^\infty ds \rho_1(s) {q^{\hf s^2+\hf
a^2}_o\o\eta(q_o)^2} }
where $\rho_0$ and $\rho_1$ are given by
\eqn\rhodef{\eqalign{
\rho_0(s)&=\int_{-\infty}^\infty dp_x{\cos(2\pi p_xs)\cosh({2\pi
p_x\o Q})\o 2\sinh(\pi p_xQ) \sinh({2\pi p_x\o Q})} \cr
\rho_1(s)&=\int_{-\infty}^\infty dp_x{\cos(2\pi p_xs)\o 2\sinh(\pi
p_xQ) \sinh({2\pi p_x\o Q})} }}
$\rho_0(s)$ is interpreted as
the spectral density of open strings connecting the same side of the
hairpin, and $\rho_1(s)$ is the spectral density of the open strings
connecting opposite sides of the hairpin
brane. Note that the open string spectrum
in $I_1$ has a gap $\hf a^2$ coming from the tension
of open string stretched between two sides of the hairpin.

\subsec{Open String Production}

As mentioned above, the Euclidean hairpin brane is T-dual to the
system with boundary potential \LukyanovBF \eqn\SdelB{
S_B=\int_{\del\Si}d\tau \,e^{-{X\o
Q}}(\si_+e^{ia\til{Y}}+\si_-e^{-ia\til{Y}}) }
Following \StromingerPC, we analyze the open string dynamics in
the mini-superspace approximation, that is we quantize the open
string massive modes accounting for the time dependence  of the
zero modes. Those massive modes are solutions of  the equation
\eqn\minieq{ \lf(\matrix{-\del_{\tilde{y}}^2-\del_x^2+m^2&e^{-{x\o
Q}+ia\tilde{y}}\cr e^{-{x\o
Q}-ia\tilde{y}}&-\del_{\tilde{y}}^2-\del_x^2+m^2}\ri)
\lf(\matrix{\psi_+\cr\psi_-}\ri)=0 } where $m$ is the open string
mass coming from the transverse momentum and the stringy
oscillators.

Solutions to \minieq\ are given by the modified Bessel functions
$I_\nu, K_\nu$. Requiring that the wavefunctions are localized at
the weak coupling end, only $K_\nu$ gives a sensible answer.
Therefore we find the solution \eqn\solpsi{
\lf(\matrix{\psi_+\cr\psi_-}\ri)=\lf(\matrix{ e^{{i\o2}
a\tilde{y}}K_\nu(2Qe^{-{x\o2Q}})\cr e^{-{i\o2}
a\tilde{y}}K_\nu(2Qe^{-{x\o2Q}})}\ri)~,\qquad \nu=Q\rt{a^2+4m^2} }
This is reminiscent of the wavefunction in the Liouville theory \GinspargIS,
in
particular the solution is not oscillatory in Lorentzian time. We
therefore find  that there is no open string production in the
mini-space approximation.

\subsec{Lorentzian Hairpin Brane}

The Lorentzian brane is naturally defined via the Wick rotation of
the position space wavefunction \NakayamaYX \eqn\psipos{\eqalign{
\til{\Psi}({\bf x})&=\int{d^2p\o(2\pi)^2}e^{-i{\bf p}\cdot {\bf
x}}\Psi({\bf p}) \cr =&{1\o2\pi aQ^2}(2\cos{y\o2a})^{-1-{2\o Q^2}}
\exp\lf[-{x\o Q}-e^{-{x\o Q}}(2\cos{y\o2a})^{-{2\o Q^2}}\ri] \cr
\uerel{y=ix^0}{\longrightarrow}~~&{1\o2\pi
aQ^2}(2\cosh{x^0\o2a})^{-1-{2\o Q^2}} \exp\lf[-{x\o Q}-e^{-{x\o
Q}}(2\cosh{x^0\o2a})^{-{2\o Q^2}}\ri] }} This is approximately
localized on the Lorentzian trajectory \eqn\Lorshape{
e^{-{Q\o2}X}=2 \cosh{X^0\o2a} }where we chose coordinates to
simplify the form of the trajectory.

By an inverse Fourier transform, the momentum space wavefunction for
the Lorentzian hairpin brane is given by \eqn\mompsi{\eqalign{
\Psi(p,\om)&=\int dx^0\,dx\,e^{i\om x^0+ipx}\til{\Psi}(x^0,x) \cr
&={1\o2\pi}{\Ga(1-ipQ)\o\Ga(1+i{2p\o Q})} \,\Ga\lf(\hf+i({p\o
Q}+a\om)\ri)\Ga\lf(\hf+i({p\o Q}-a\om)\ri) }}

Interestingly, the wavefunction \psipos\ can be written as the
smearing of the trajectory \Lorshape\ over its modulus (expressing
the maximal location of the trajectory, or equivalently the
initial velocity at some point in time). Indeed
\eqn\smearpsi{\eqalign{
 \Psi(x,x^0)&={1\o4\pi
aQ^2\cosh{x^0\o2a}}\int_{-\infty}^\infty ds\, w(s)\,
\cob\lf(x-s+{2\o Q}\log\Big(2\cosh{x^0\o2a}\Big)\ri)  \cr &={1\o4\pi
aQ}\int_{-\infty}^\infty ds\, w(s)\, \cob\lf(e^{-{Q\o2}(x-s)}-
2\cosh{x^0\o2a}\ri) }} where the smearing function $w(s)$
 is given by \foot{This can
be shown using the expression $\Gamma(z) = \int_{-\infty}^{\infty}
dt\, \exp(-zt-e^{-t})$.}
\eqn\rhos{ w(s)=\exp\lf(-{s\o Q}-e^{-{s\o Q}}\ri) } In
other words, the wavefunction is a superposition of the trajectory
$e^{-{Q\o 2}(x-s)}=2\cosh{x^0\o2a}$ with the weight $w(s)$. We can
see that $w(s)$ is localized around $s=0$ \eqn\rhosgauss{ w(s)\sim
\exp\lf(-{s^2\o2Q^2}\ri) ~~{\rm for} ~~s\sim 0} Note that the width
of distribution is $\lap s=Q$ \KutasovRR\ and $w(s)$ decays to zero
as $s\riya\pm\infty$. This smearing comes from the factor
$\Ga(1-ip_xQ)$ in the momentum space wavefunction \mompsi, which in
turn is identified in Euclidean space as the effect of open string
winding tachyon.

We note that if we interpret the wavefunction $w(s)=\exp\lf(-{s\o
Q}-e^{-{s\o Q}}\ri)$ as the profile of the zero mode $s$, the
profile of the D-brane itself is given by\foot{In a classical
field theory the zero mode profile is a derivative of the
classical solution.} $e^{-e^{-{s\o Q}}}$. This has the
semiclassical form $e^{-I(s)}$, with $I(s)=e^{-{s\o Q}}$ being the
action of the screening charge $e^{-{X\o Q}}$.

\fig{Stringy smearing of the D-brane trajectory. The classical
trajectory $e^{-{Q\o2}x}=2\cosh{t\o2a}$ is smeared by the weight
function $w(s)$ with width $Q$. }{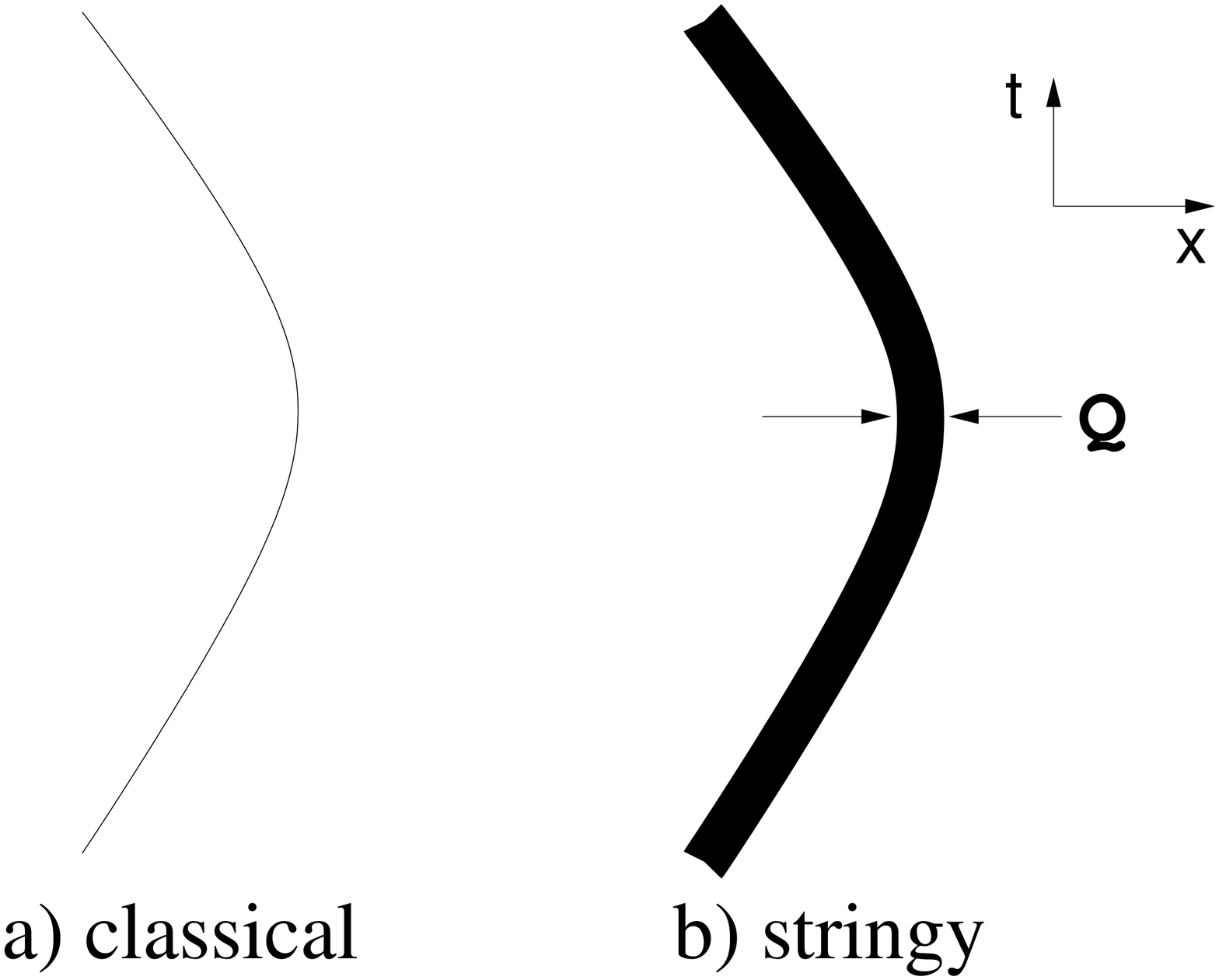}{5cm}

\subsec{The Closed String Field}

  As discussed in \MaloneyCK, the
classical closed string field $|C\ket$ sourced by the boundary state
$|B\ket$  is found by solving the equation
$(L_0+\b{L}_0)|C\ket=|B\ket$, imposing retarded boundary conditions.
As is familiar with discussions of accelerated trajectories, there
is an effective Rindler  horizon in spacetime -  causality prevents
any signal from reaching a part of the spacetime. In our case that
part includes all of the weakly coupled asymptotic region.

To see that effect we concentrate on the lowest mode of the closed
string field, namely the closed string tachyon. Neglecting for the
moment the stringy smearing of the trajectory, the Lorentzian
boundary state can be written in lightcone coordinates. Defining
$p_{\pm}= {p\o Q}  \pm a \om$ we find in the $Q \rightarrow 0$ limit
\eqn \bs{\Psi(p_+,p_-) =  {1\o {2\pi}}\,{\Gamma({1\o2} + i
p_+)\,\Gamma({1\o2} + i p_-) \o \Gamma(1 + ip_+ + i p_-)}}

This localizes the position space wave function on the trajectory
$e^{x^+} +e^{x^-}=1$, where $x^\pm$ are the positions conjugate to
the lightcone momenta $p_\mp$ respectively, namely $x^\pm = {Q \o
2} x \mp {x_0\o {2a}}$. This is of course the trajectory
\Lorshape\ written in lightcone coordinates. Note that this
implies that both $x^\pm$ are negative, so that the trajectory is
localized on one quadrant of the two dimensional Minkowski space,
as shown in figure 2.

\fig{The trajectory in the $Q\riya0$ limit is contained in one
quadrant $x^+<0,x^-<0$. Therefore, the closed string radiation is
observed only in the shaded region, {\it i.e.} the future
lightcone of the trajectory. This picture is completely changed
when including stringy effects.}{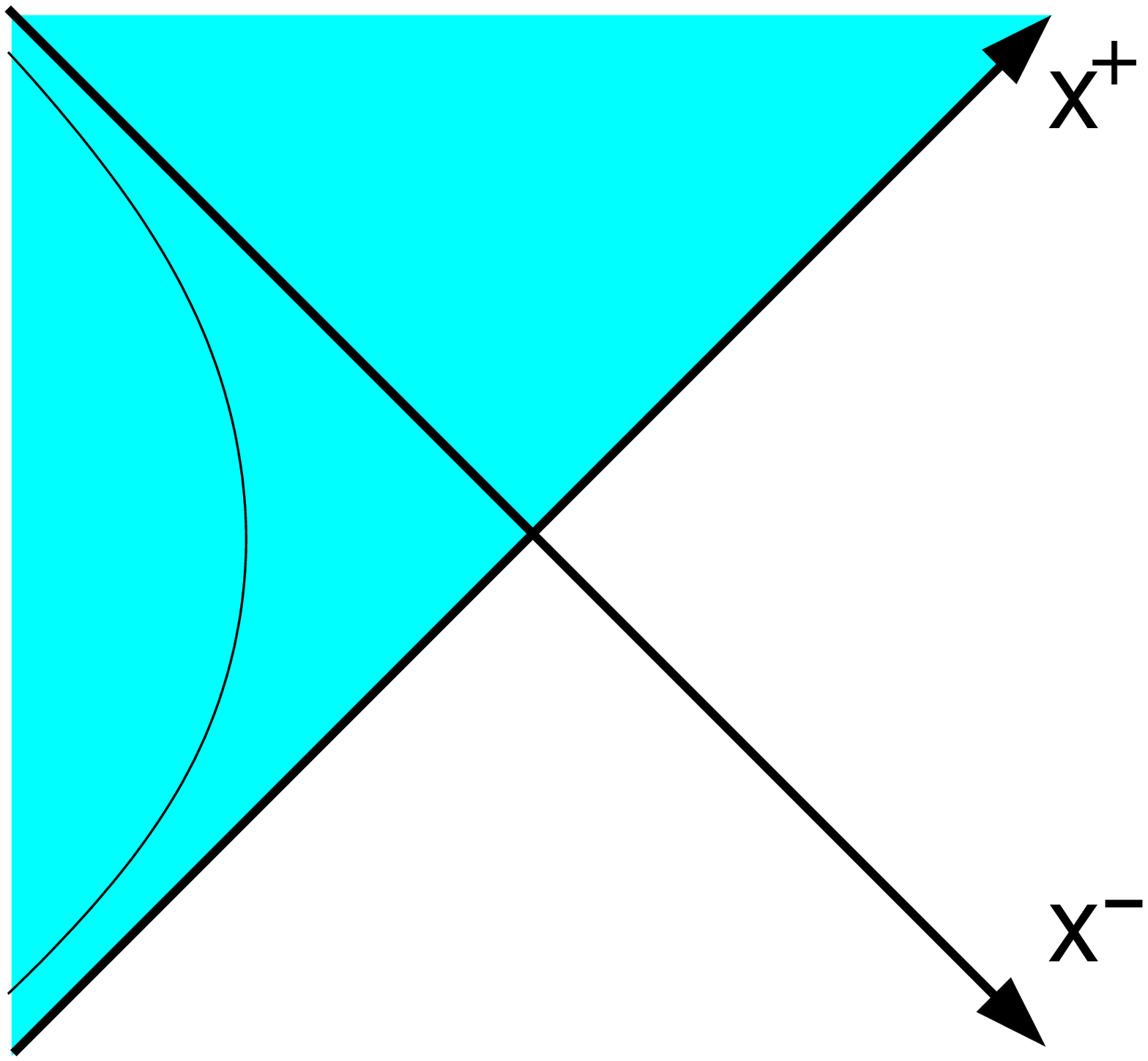}{4cm}

The retarded propagator for a two-dimensional massless field is \eqn
\ret{ G_R(p_+,p_-) \sim {1 \o {(p_- -i \epsilon) (p_+ +
i\epsilon)}}} where we omit an overall constant and use the fact
that in the $Q\rightarrow 0$ limit $a \rightarrow {1\o Q}$.
Therefore the massless closed string field generated by our
trajectory is proportional to \eqn \closed { T(x^+,x^-) \sim \int
dp_- \, dp_+ \exp(-i p_-x^+ -i p_+ x^-)\,{1 \o {(p_- -i \epsilon)
(p_+ + i\epsilon)}}\, {\Gamma({1\o2} + ip_+)\,\Gamma({1\o2} +i p_-)
\o \Gamma(1 + ip_+ + ip_-)\, } }

The integral is easily evaluated, as are similar integrals for the
massive closed string fields. We will not be interested in the
detailed structure of the closed string field, rather just in its
causal structure. To this end we look at the poles of the
integrand of \closed\  and note that they appear only for ${\rm
Im} (p_-) > 0$ (whereas there are poles in both upper and lower
half planes of $p_+$, when taking into account the poles in the
gamma functions). Therefore it is clear that the closed string
field vanishes for $x^- >0$. This fits with our expectations from
causality, as drawn in figure 2.

The effect of stringy corrections to the boundary state smear the
trajectory as described above, and therefore it is clear that the
resulting closed string field will be everywhere non-vanishing. This
can be seen by noticing that the additional factor in the boundary
state $\Gamma(1- i Q^2 (p_++p_-))$ has poles both in the upper and
lower half planes of $p_-$. Indeed, an estimate of the profile of
the closed string field at large positive $x^-$, based on the first
pole in the lower half plane, gives $T\sim e^{-{2\o Q} x^-}$.

Therefore we find that the closed string field has exponential tail
in a region naively forbidden by causality, resulting from stringy
smearing of the trajectory of the D-brane.


\subsec{Closed String  Radiation}

Next we wish to calculate the imaginary part of annulus amplitude
between two hairpin branes. We start with the $(X^0,X)$ part
\eqn\ZXzero{Z^{X^0X}= \int dp\,
d\om|\,\Psi(p,\om)|^2\,\eta(q_c)^{-2}\,q_c^{\hf(p^2-\om^2)} }
Following \KarczmarekXM\ we rewrite this in terms of the
 characters in the open string channel. Using \formeta\ we obtain  \eqn\openxzeroo{\eqalign{
Z^{X^0X}&=\int dk\,dE\,dp\,d\om|\,\Psi(p,\om)|^2\,\cos({2\pi
kp})\,\cos(2\pi E\om)\, \eta(q_o)^{-2}\,q_o^{\hf(k^2-E^2)} \cr
&={1\o 2aQ}\int dk\,dE{\sinh{2\pi E\o aQ^2}\o\sinh{\pi E\o a}
(\cosh{2\pi k\o Q}+\cosh{2\pi E\o
aQ^2})}\,\eta(q_o)^{-2}\,q_o^{\hf(k^2-E^2)} }} where $q_o$ is the
open string modulus, and $E,k$ are respectively the energy and
momentum in the open string channel (roughly the T-duals of $\om,
p$).

 As
the integral over $q_0$ is ill-defined, due to the non-unitary
factor $X^0$ in the CFT, we need to make sense of this expression.
 We do that by Euclidean rotation of the variable $E$, resulting
 in
 \eqn\ZXzero{ Z^{X^0X}={1\o 2aQ}\int dk\,dE{\sin{2\pi E\o aQ^2}\o\sin{\pi E\o a}
(\cosh{2\pi k\o Q}+\cos{2\pi E\o
aQ^2})}\,\eta(q_o)^{-2}\,q_o^{\hf(k^2+E^2)} } where $E$ is now the
Euclidean frequency, and it is integrated over a Feynman contour,
which includes the poles of the expression for positive values of
$E$.

In order to obtain a critical string background we couple the
$(X^0,X)$ CFT to $D$ free bosons (or more generally some compact
CFT with central charge $D$) such that \eqn\Dtwosix{ D+2+3Q^2=26.
}
The contribution from $D$ free boson and the $bc$ ghosts is
\eqn\bcX{\eqalign{ Z^{X^i+bc}&=\int
d^{D-p}k_{\perp}\eta(q_c)^{2-D}q_c^{\hf k_{\perp}^2} \cr
&=t_o^{1-{p\o2}}\eta(q_o)^{2-D} }} Here we assumed that the $p$
bosons satisfy the Neumann boundary condition and $D-p$ bosons
satisfy the Dirichlet condition. Putting it all together the total
annulus amplitude is given by \eqn\totalZ{ {\cal Z}={1\o
2aQ}\int_0^\infty{dt_o\o t_o}t_o^{-{p\o2}}\eta(q_o)^{-D}
\int_{-\infty}^\infty dk\int_{-\infty}^\infty dE {\sin{2\pi E\o
aQ^2}\o\sin{\pi E\o a} (\cosh{2\pi k\o Q}+\cos{2\pi E\o
aQ^2})}\,q_o^{\hf(k^2+E^2)} }

In the form \totalZ\ the partition function is well-defined and one
can extract its imaginary part. Indeed, due to the poles at
$E=na~(n\in{\Bbb Z})$ on the real $E$-axis, and using the Feynman
contour, ${\cal Z}$ acquires the imaginary part  \eqn\imcalZ{ {\rm
Im}{\cal Z}={1\o 2Q}\int_0^\infty{dt_o\o
t_o}t_o^{-{p\o2}}\eta(q_o)^{-D} \int_{-\infty}^\infty
dk\sum_{n=1}^\infty {(-1)^n\sin{2\pi n\o Q^2}\o \cosh{2\pi k\o
Q}+\cos{2\pi n\o Q^2}}q_o^{\hf(k^2+n^2a^2)} } The imaginary part is
to be interpreted as the total radiation of closed strings. We now
discuss some of its features.

First, it is interesting to check whether the total radiation is
finite. To that end we look at the open string IR limit
$t_o\riya\infty$, the amplitude is then dominated by the $n=1$ pole,
and behaves as \eqn\asymIR{ {\rm Im}{\cal Z}=-{1\o
Q}\int^\infty{dt_o\o t_o}t_o^{-{p\o2}} \int_{-\infty}^\infty dk
{\sin{2\pi \o Q^2}\o \cosh{2\pi k\o Q}+\cos{2\pi \o Q^2}}e^{-\pi
t_o( k^2+ a^2-{D\o12})}+\cdots } We note that the decay rate is
exponentially suppressed with an exponent which is positive definite
\eqn\aDexp{ a^2-{D\o12}=\lf(1+{1\o Q^2}\ri)-{24-3Q^2\o12}= \lf({1\o
Q}-{Q\o2}\ri)^2 } This exponential suppression can be simply
understood in the open string channel, since the ground state  of
the stretched open strings is massive, as in \KutasovRR. To see
this, let us look at the action of open string tachyon at the weak
coupling end\eqn\Ltach{ {\cal
L}=-e^{{Q\o2}x}\Big[(\del_xT)^2+m(x)^2T^2\Big] } where $m(x)^2$ is
the position dependent mass \eqn\mxmass{
m(x)^2=-\hf+{1\o\pi^2}a^2\arccos^2(e^{-{Qx\o 2}}) } The first term is
the usual tachyon mass and the second term is the tension of open
string stretched between two sides of the Euclidean hairpin brane.
In terms of the canonically normalized field $\til{T}=e^{{Q\o4}x}T$,
the effective mass becomes \eqn\tilmass{
\til{m}(x)^2=m(x)^2+{Q^2\o16} } The tachyon mass at the weak
coupling side is \eqn\massiny{ \lim_{x\riya\infty}
\til{m}(x)=-\hf+{a^2\o4}+{Q^2\o16}=\qu\lf({1\o Q}-{Q\o2}\ri)^2 }
This agrees with \aDexp\ up to a normalization of momentum.

We conclude then that the closed string radiation is finite, and
becomes infinite exactly at the point where the Lorentzian hairpin
brane becomes non-normalizable ($Q= \sqrt{2}$). This is reminiscent
of the discussion of the rolling tachyon in \KarczmarekXM, and contradicts
the results in \NakayamaYX\foot{In addition to discussing a slightly
different model, the analysis in \NakayamaYX\ uses saddle point
approximation which becomes invalid at large mass levels.}.

As the decay rate is finite we expect it to be dominated by
radiation to few of the lowest mass levels of the closed string. As
the characters are exactly known we could calculate the emission
rate of any closed string field. For example, in order to extract
the contribution of the light closed string modes, we take  the
closed string IR limit $t_o\riya0$, the amplitude behaves as
\eqn\closedIR{\eqalign{ {\rm Im}{\cal Z}=&{1\o2 Q}\int_0{dt_o\o
t_o}t_o^{{D-p\o2}}\lf( e^{2\pi
D\o24t_o}+De^{2\pi(D-24)\o24t_o}+\cdots\ri) \cr &\times
\int_{-\infty}^\infty dk\sum_{n\in{\Bbb Z}} {(-1)^n\sin{2\pi n\o
Q^2}\o \cosh{2\pi k\o Q}+\cos{2\pi n\o Q^2}}q_o^{\hf(k^2+n^2a^2)} }}
The first term is the closed string tachyon contribution. The second
term is exponentially suppressed because $D-24=-3Q^2<0$. This is a
manifestation of the fact that the graviton in the linear dilaton
background is screened and is effectively massive.

Finally, we  discuss the details of the radiation as function of
time or frequency. To this end  we can calculate the moments of the
radiation, or equivalently the generating function for those
moments.  That generating function is obtained by inserting a factor
$e^{i\om t}$ in \ZXzero. Repeating the steps above gives an
expression of the form \totalZ\ with the replacement \eqn\sintoft{
{1\o\sin{\pi E\o a}}\riya f(t,E)= {2\sin{\pi E\o a}\cosh{t\o2a}\o
\cosh{t\o a}-\cos{2\pi E\o a}} } The Fourier transform of $f(t,E)$
is \eqn\tilf{ \til{f}(\om,E)=\int dt e^{-i\om t}f(t,E) =2\pi
a{e^{-2\pi\om E}\o 1+e^{2\pi a\om}} } Therefore  the annulus
amplitude is written as \eqn\Zinomint{ {\cal Z}(t)=\int d\om e^{i\om
t}F(\om) } \eqn\Fomdef{ F(\om)={1\o 2aQ}\int_0^\infty{dt_o\o
t_o}t_o^{-{p\o2}}\eta(q_o)^{-D} \int_{-\infty}^\infty
dk\int_{-\infty}^\infty dE {\til{f}(\om, E)\sin{2\pi E\o aQ^2}\o
\cosh{2\pi k\o Q}+\cos{2\pi E\o aQ^2}}q_o^{\hf(k^2+E^2)} }

We note that the expression \tilf\ is a Boltzmann factor written
in terms of the open string (Euclidean) frequency. This suggests
that the system, when formulated in the open string variables,  is
thermal with temperature   \eqn\temp{ T={1\o2\pi a} } which is the
expected Unruh temperature of the accelerated brane. As another
piece of evidence we note that even though the system has no
Euclidean time periodicity in the original variables, it does have
such periodicity in the dual variables \SdelB. We note that this
temperature is lower than the closed string Hagedorn temperature
for the physical range of the parameter $Q$.

\newsec{D-Branes on the Two Dimensional Black Hole}

In this section we discuss the hairpin brane in the two-dimensional
black hole background. In addition to discussing the features of
the brane outside the horizon, previously also discussed in
\NakayamaPK, we suggest an analytic continuation of the boundary
state emphasizing the geometrical trajectory of the particle. With
this analytic continuation we find a surprising result: the
portion of the trajectory inside the horizon looks to an
asymptotic observer as a ghost D-brane \OkudaFB.

\subsec{D1-Branes on the Euclidean Cigar}

The Euclidean cigar $SL(2,{\Bbb R})/U(1)$ CFT is described by the
background metric and dilaton \eqn\cigargphi{
ds^2=k(d\rho^2+\tanh^2\rho d\th^2),\quad \Phi=-\log\cosh\rho } and
the central charge is \eqn\ccigar{ c=2+6b^2, \quad b^2={1\o k-2} }
D-branes in the background were constructed in \RibaultSS. We are
interested in the D1-brane on the cigar whose
 disk one-point function of primary operator $\phi_n^p$ is given by
\eqn\Donecigar{ \Psi_{(\rho_0,\th_0)}(p,n)={\Ga(ip)\Ga(1+ipb^2)
\o\Ga(\hf+i{p\o2}+{n\o2})\Ga(\hf+i{p\o2}-{n\o2})}e^{in\th_0}
(e^{-ip\rho_0}+(-1)^ne^{ip\rho_0}) } where $\rho_0(>0)$ and $\th_0$
are constant parameters specifying the classical shape of D1-brane
\eqn\trajEuc{ \cos(\th-\th_0)\sinh\rho=\sinh\rho_0 }

Following \NakayamaPK, we rewrite \Donecigar\ as the integral of
mini-superspace closed string wavefunctions over the trajectories.
In the mini-superspace approximation, the eigenfunction of the
target space Laplacian $\lap={1\o e^{-2\Phi}\rt{g}}
\del_i(e^{-2\Phi}\rt{g}g^{ij}\del_j)$ is given by \eqn\eigenphi{
\phi_n^p=\phi_{L,n}^p+{\cal R}_0(p,n)\phi_{R,n}^p } where
$\phi_{L,n}^p$ and $\phi_{R,n}^p$ are the left and right moving
modes \eqn\LRmodes{\eqalign{
\phi_{L,n}^p&=e^{in\th}(\sinh\rho)^{-1-ip}F\lf({1+ip+n\o2},{1+ip-n\o2},1+ip,
-{1\o\sinh^2\rho}\ri) \cr
\phi_{R,n}^p&=e^{in\th}(\sinh\rho)^{-1+ip}F\lf({1-ip-n\o2},{1-ip+n\o2},1-ip,
-{1\o\sinh^2\rho}\ri) }} and ${\cal R}_0(p,n)$ in \eigenphi\ is the
mini-superspace reflection amplitude \eqn\miniref{ {\cal
R}_0(p,n)={\Ga(ip)\Ga({1-ip+n\o2})^2\o\Ga(-ip)\Ga({1+ip+n\o2})^2 } }
which related spacetime left and right movers (as the coordinate
$\rho$ is semi-infinite).
Note that $\phi^p_n$ and $\phi^{-p}_n$ are related by the
reflection
\eqn\refeuc{
\phi^p_n={\cal R}_0(p,n)\phi^{-p}_n.
}
Therefore, the boundary state is expanded by the Ishibashi states
$\bra\bra\phi^p_n|$ with $p>0$
\eqn\Bexpeuc{
\bra B|=\int_0^\infty dp\sum_{n\in\Z}\Psi_{(\rho_0,\th_0)}(p,n)
\bra\bra\phi^p_n|
}

In \NakayamaPK, it was shown that the disk one-point function for
the D1-brane \Donecigar\ is written as the integral of mini-superspace
wavefunction $\phi_n^p(\rho,\th)$
over the classical trajectory \trajEuc \eqn\doneint{
\Psi_{(\rho_0,\th_0)}(p,n)=\Ga(1+ipb^2)\int d\mu\,
\cob\Big(\cos(\th-\th_0)\sinh\rho-\sinh\rho_0\Big)\phi_n^p(\rho,\th)
} where the measure $d\mu$ is given by \eqn\dmudef{ \int d\mu=\int
e^{-2\Phi}\rt{g}d\rho d\th= \int_0^\infty \sinh\rho
d(\sinh\rho)\int_{0}^{2\pi}d\th } This is analogous to Fourier
transform of the boundary state in the hairpin case. As in the
hairpin case, the factor $\Ga(1+ipb^2)$ in \doneint\
 has the
effect of smearing the trajectory and it can be attributed to the
condensate of open string tachyon. It is curious that in both
cases the sole effect of $\alpha'$ corrections is summarized in
rigid smearing of the classical trajectories.

We can compute the annulus amplitude between two D1-branes as in
the hairpin case: \eqn\anuDoneci{\eqalign{
&Z(\rho_0,\th_0|\rho_0',\th_0')\cr =&\int_0^\infty
dp\sum_{n\in{\Bbb Z}}{q_c^{{b^2p^2\o4}+{n^2\o4k}}
\o\eta(q_c)^2}\Psi_{(\rho_0,\th_0)}(p,n)
\Psi_{(\rho_0',\th_0')}^*(p,n) \cr =&\int
ds{q^{{b^2s^2\o4}}_o\o\eta(q_o)^2}\lf[d_0(s|\rho_0,\rho_0')\sum_{w\in{\Bbb
Z}}
q^{k(w+{\th_0-\th_0'\o2\pi})^2}_o+d_1(s|\rho_0,\rho_0')\sum_{w\in{\Bbb
Z}+\hf} q^{k(w+{\th_0-\th_0'\o2\pi})^2}_o\ri] }} where $d_{0}$ and
$d_1$  are the open string density of states defined by
\eqn\rhoonetwo{\eqalign{ d_0(s|\rho_0,\rho_0')&=\int_0^\infty
dp\cos(\pi ps){\cosh({\pi p\o2})\cos(p\rho_0)\cos(p\rho_0')
\o\sinh({\pi p\o2})\sinh(\pi b^2p)} \cr
d_1(s|\rho_0,\rho_0')&=\int_0^\infty dp\cos(\pi ps){\sinh({\pi
p\o2})\sin(p\rho_0)\sin(p\rho_0') \o\cosh({\pi p\o2})\sinh(\pi
b^2p)} }} We can interpret the $d_0$ term in \anuDoneci\ as the
open string stretched between the same side of D1-brane, while
$d_1$ term is the contribution of the open string stretched
between the opposite sides of D1-brane. As noted in \LukyanovNJ,
the structure of the annulus amplitude for the D1-branes on the
cigar is very similar to the annulus amplitude of the Euclidean
hairpin brane.

\subsec{The Lorentzian Black Hole}

The two-dimensional Lorentzian black hole is obtained from the
Euclidean one \cigargphi\ by a Wick rotation $\th=it$ \WittenYR
\eqn\Lorentzgphi{ ds^2=k(d\rho^2-\tanh^2\rho dt^2),\quad
\Phi=-\log\cosh\rho. } In this Schwarzschild coordinate system,
the trajectory of D0-brane can be  obtained by the Wick rotation
of the Euclidean D1-brane \trajEuc \eqn\trajrhot{
\cosh(t-t_0)\sinh\rho=\sinh\rho_0 } However, in order to discuss
the global properties of the geometry, and the full trajectory of
the D-brane, it is useful to introduce the  Kruskal coordinate
\eqn\uvinrhot{ u=\sinh\rho e^t,\quad v=\sinh\rho e^{-t}. } In
terms of these coordinates, the black hole geometry is given by
\eqn\uvmet{ ds^2=2k{dudv\o1+uv},\quad e^{2\Phi}={1\o1+uv} }

The $uv$-plane is divided into various regions due to the presence
of the horizon at $uv=0$ (see fig. 3.). In particular,
\eqn\regionuv{\eqalign{ {\rm outside~the~horizon}:&~uv>0 \cr {\rm
behind~the~horizon}:&~uv<0. }} There is a spacelike singularity at
$uv=-1$. Note that in string theory it may be  sensible to
include the region beyond the singularity $uv<-1$.  In particular
we will find that the boundary state of the D-particle does not
see the singularity at this level of approximation.

\fig{The two-dimensional black hole in the Kruskal coordinate. The
horizon is at $uv=0$ and the singularity is at $uv=-1$. The
trajectory of D0-brane in this coordinate is a straight line,
which consists of three parts ($a,b$ and $c$  in the figure).
Correspondingly, the disk one-point function has two parts, $
\Psi_{\rm outside}=b, \Psi_{\rm behind}=a+c. $ The classical
trajectory is smeared by the open string tachyon.}
{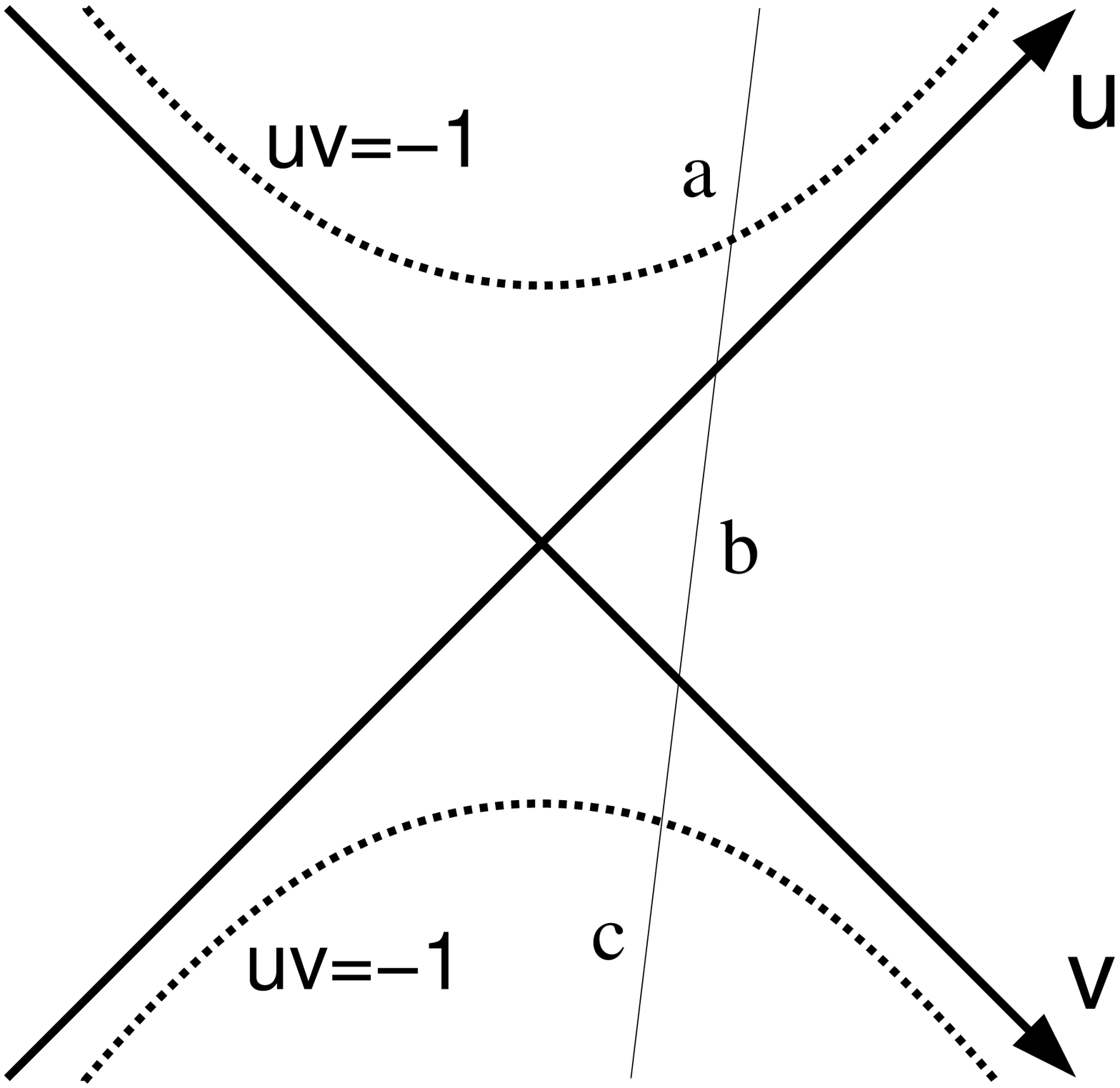}{5cm}

The closed string wavefunctions in the Kruskal coordinate are
given by the hypergeometric functions \eqn\LRinuv{\eqalign{
R^p_\om&=u^{-\nu_-}v^{-\nu_+}F\Big(\nu_-,\nu_+,\nu_-+\nu_+;-{1\o
uv}\Big)\cr
L^p_\om&=u^{-\b{\nu}_+}v^{-\b{\nu}_-}F\Big(\b{\nu}_-,\b{\nu}_+,
\b{\nu}_-+\b{\nu}_+;-{1\o uv}\Big) }} Here we defined \eqn\nupm{
\nu_{\pm}=\hf-i{p\pm\om\o2} } As discussed in \DijkgraafBA, these
eigenfunctions \LRinuv\ can be  analytically continued to the
entire $uv$-plane. This basis of functions satisfy natural
boundary conditions at infinity, namely they represent purely left
or right movers at large radial distances.
In contrast to the Euclidean case, mode functions with $p>0$ and
$p<0$ are independent. Since $L^p_\om$ and $R^p_\om$ are related
by the sign flip of momentum \eqn\refLR{ L^{-p}_\om=R^p_\om } we
can take the basis of Ishibashi states as \eqn\ishiLR{
\Big\{|L^p_\om\ket\ket,|R^p_\om\ket\ket\Big\}_{p>0,\om\in{\Bbb R}}
=\Big\{|L^p_\om\ket\ket\Big\}_{p,\om\in{\Bbb R}}. }

Note that the reflection relation \refeuc\ satisfied by the
Euclidean modes amounts to imposing Hartle-Hawking boundary
conditions at the outer horizon\foot{It is interesting that those
boundary conditions receive stringy corrections.}, which is a
natural set of boundary condition needed when discussing the physics
outside the horizon. As one of our goal is to understand the
D-particle in the full geometry, we do not impose any such
reflection relations and leave the modes independent.

Another natural set of eigenfunctions $U^p_\om,V^p_\om$ is
characterized by boundary conditions at the horizon, they are
purely incoming or outgoing modes at the horizon. This basis  is
related to $L^p_\om, R^p_\om$ by \eqn\UVdef{\eqalign{
U^p_\om&=L^p_\om+{\cal R}_0(p,\om)R^p_\om \cr
V^p_\om&=R^p_\om+{\cal R}_0^*(p,\om)L^p_\om }} where ${\cal R}_0$
is the mini-superspace reflection amplitude \eqn\Rlorentz{ {\cal
R}(p,\om)=-{B(\nu_+,\nu_-)\o
B(\b{\nu}_+,\b{\nu}_-)}S(p,\om),\qquad
S(p,\om)={\cosh(\pi{p-\om\o2})\o\cosh(\pi{p+\om\o2})} } and
$B(a,b)={\Ga(a)\Ga(b)\o\Ga(a+b)}$ denotes the Euler beta-function.

We note that these wavefunctions all diverge logarithmically near
the singularity at $uv=-1$. We will find that this divergence
disappears when the closed string modes are combined in the wave
packet corresponding to the wavefunction of the D-particle.

\subsec{D-particles and ghost particles in the Lorentzian Black Hole}

The  D0-brane trajectory  \trajrhot\ becomes, when expressed in
Kruskal coordinates, a straight line on the $uv$-plane \refs{\SenYV,\YogendranDM}
\eqn\trajuv{ ue^{-t_0}+ve^{t_0}=2\sinh\rho_0. } This can be
understood by the fact that the effective metric seen by the brane
is flat: $ds^2_{D0}=dudv$, the conformal factor of the metric and
the dilaton precisely cancel each other. In particular, before
including stringy or quantum corrections there is no special
status to the singularity at $uv=-1$.

In order to  analytically continue the   Euclidean boundary state
\doneint, we need to  define the disk one-point function of
$L^p_\om$ for the D0-brane in the black hole background. Since there
are more such modes in the Lorentzian black hole background than the
Euclidean section (as modes with positive and negative momenta are
unrelated), there is some freedom in determining these one point
function.

Our prescription relies on the geometric interpretation of the
trajectory: the one point functions are obtained by integrating the
mini-superspace eigenfunction $L^p_\om(u,v)$ along the classical
trajectory. This in effect imposing "transparent horizon" boundary
conditions, in other words the D-brane propagation through the
horizon is smooth: \eqn\psifdef{ \Psi^L(p,\om)=\Ga(1+ ipb^2)\int
d\mu\, \cob\Big(ue^{t_0}+ve^{-t_0} -2\sinh\rho_0\Big)L^p_\om(u,v) }
where the measure $d\mu$ is \eqn\measureuv{ \int d\mu=\int
e^{-2\Phi}\rt{-g}dudv= \int du dv } Note that the measure in the
$uv$-coordinate is flat, thus there is no singularity in the measure
factor. This is related to the cancelation between the dilaton and
conformal factor in the action for the D-particle.


In  \psifdef\ we have to specify the integration region of $u,v$
variables. There are two natural choices of integration region: we
can integrate over  the whole $uv$-plane, or just over the region
outside the horizon $uv>0$. In the former case   the eigenfunction
$L^p_\om(u,v)$ should be extended to the entire $uv$-plane as in
\DijkgraafBA. Explicitly, we extend $L^p_\om(u,v)$ from one causal
patch $u,v>0$ to the entire $uv$-plane as follows: \eqn\Lext{
L^p_\om(u,v)=\lf\{ \eqalign{&e^{-\pi
i\b{\nu}_+}(-u)^{-\b{\nu}_+}v^{-\b{\nu}_-}F(\b{\nu}_+,\b{\nu}_-,\b{\nu}_++\b{\nu}_-;-1/uv)\cr
&e^{\pi
i\b{\nu}_-}u^{-\b{\nu}_+}(-v)^{-\b{\nu}_-}F(\b{\nu}_+,\b{\nu}_-,\b{\nu}_++\b{\nu}_-;-1/uv)\cr
&e^{\pi i(\b{\nu}_--\b{\nu}_+)}(-u)^{-\b{\nu}_+}
(-v)^{-\b{\nu}_-}F(\b{\nu}_+,\b{\nu}_-,\b{\nu}_++\b{\nu}_-;-1/uv)}
\quad\eqalign{&u<0,v>0 \cr &u>0,v<0\cr &u<0,v<0}\ri. } Our phase
convention is such that the Schwarzschild time $t$ is shifted by
$-\pi i$ when crossing a horizon. The extension of $R^p_\om$ is
obtained similarly by requiring \refLR. Note that this is the
prescription that maps naturally to real time thermal field theory.

Let us now consider the contribution to the boundary state  from the
regions outside the horizon and  behind the horizon separately
\eqn\BDzero{\eqalign{ \bra B|\equiv\bra B|_{\rm outside}
&=\int_{-\infty}^\infty {dp\o2\pi}\int_{-\infty}^\infty{d\om\o2\pi}
\bra\bra L^p_\om| \Psi^L(p,\om)_{\rm outside}\cr \bra B|_{\rm
behind} &=\int_{-\infty}^\infty
{dp\o2\pi}\int_{-\infty}^\infty{d\om\o2\pi} \bra\bra
L^p_\om|\Psi^L(p,\om)_{\rm behind} }} where the disk one-point
functions of $L^p_\om$ are given by \eqn\defbehind{\eqalign{
\Psi^L(p,\om)_{\rm outside}&=\Ga(1+ ipb^2) \int_{uv>0} dudv\,
\cob\Big(ue^{t_0}+ve^{-t_0} -2\sinh\rho_0\Big)L^p_\om(u,v) \cr
\Psi^L(p,\om)_{\rm behind}&=\Ga(1+ ipb^2)\int_{uv<0} dudv\,
\cob\Big(ue^{t_0}+ve^{-t_0} -2\sinh\rho_0\Big)L^p_\om(u,v) }} Using
the explicit form of the eigenfunctions \LRinuv, we find a
remarkable property\foot{ This conclusion does not change if we
choose $t\riya t+\pi i$ instead of $t\riya t-\pi i$ for the crossing
of horizon.} \eqn\psiinout{ \Psi^L(p,\om)_{\rm
behind}=-\Psi^L(p,\om)_{\rm outside} } It should be emphasized that
the integral behind the horizon is finite even though it includes
the contribution near and behind the  singularity. The finiteness of
the contribution from singularity can be seen as follows: while the
function $L^p_\om$ blows up as $\log(1+uv)$ near $uv=-1$, its
integral behaves  like $(1+uv)\log(1+uv)$ which is finite near
$uv=-1$. This is consistent with the classical picture that the
D0-brane does not see the singularity. As   the stringy corrections
amount to smearing over the classical trajectory, we find that
inclusion of these $\alpha'$ corrections does not change the
picture.

The striking relation \psiinout\ suggests that the portion of the
D-brane behind the horizon (and the singularity) can be interpreted
by an asymptotic observer as
 a ``ghost D-brane'' introduced in
\OkudaFB. Namely, the D-brane behind the horizon is the same as the
D-brane outside the horizon with an overall minus sign \eqn\Bbehind{
|B\ket_{\rm behind}=-|B\ket } Note that in the $c=1$ matrix model
context such ghost D-brane is identified as the hole state
\refs{\DouglasUP,\GaiottoYF}. Therefore, we propose to identify the
D-brane outside the horizon as a particle and the D-brane behind the
horizon as a hole. This suggests a relation to the black hole
complementarity \SusskindIF. It is also reminiscent of  picturing
the Hawking radiation as pair (particle-hole) creation near the
horizon.

 We note also that once the D-particle crosses the horizon
it becomes visible to the second asymptotic region of the geometry,
which is related to the first one by complex time shift and time
reversal \FidkowskiNF. Excitations of this boundary appear ghostlike
precisely as they appear in  real time thermal field theory
\refs{\IsraelUR,\MaldacenaKR}.

 Since the dilaton diverges near the singularity we
do expect quantum corrections to reintroduce some singular behavior.
We expect therefore corrections to the exact relations \psiinout,
but those corrections will be small for suitable wavepackets. In
addition, the fact that the segment inside the horizon is ghostlike,
is not going to be modified, even in the absence of the relation
\psiinout.

The relation \psiinout\ imposes, via the Cardy condition, unusual
quantization condition on the open strings stretched between the
portions of the D-brane outside and inside the horizon. Indeed the
relative sign means that the ground state of such open string that
crosses the horizon is fermionic. It would be interesting to
understand the open string sector on the D-brane in more detail.

Returning to the full boundary state \psifdef, the explicit form of
the disk one-point function of $L^p_\om$ is found to be \eqn\LRpsi{
 \Psi^L(p,\om)_{\rm
outside} =e^{-i\om t_0-ip\rho_0}B(\nu_+,\nu_-)\Ga(1+ipb^2) } This
agrees with the result in \NakayamaPK\ obtained from a different
method of analytic continuation. Indeed, the portion of our D0-brane
outside the horizon corresponds to the time-symmetric brane in
\NakayamaPK.

As was the case for the hairpin brane in flat space, the factor
$\Ga(1+ ipb^2)$ can be interpreted similarly as the smearing of
the trajectory. To see this, note that \LRpsi\ is rewritten as
\foot{One is tempted to write the one-point function in the form
$$
\Psi^L(p,\om)_{\rm outside}
=\int_{-\infty}^\infty ds \,w(s)
\int_{uv>0} dudv\,\cob\Big(ue^{-t_0}+ve^{t_0}-2\sinh(\rho_0+s)\Big)L^p_\om.
$$
However, this is not correct because the $uv$-integral is proportional
to $e^{ -ip|\rho_0+s|}$, which is not analytic in $s$.
}
\eqn\smearrho{
\Psi^L(p,\om)_{\rm outside}
=\int_{-\infty}^\infty ds \,w(s)e^{-i\om t_0-ip(\rho_0+s)}B(\nu_+,\nu_-)
}
where the weight $w(s)$ has the same form as the hairpin case \rhos
\eqn\wsdzero{
w(s)={1\o b^2}\exp\lf(-{s\o b^2}-e^{-{s\o b^2}}\ri)
={d\o ds}\exp\lf(-e^{-{s\o b^2}}\ri)
}
In other words, the parameter $\rho_0$ in the classical trajectory
fluctuates with width $\lap\rho_0=b^2$.





Note that in our definition the boundary state $|B\ket$ is
independent of the choice of basis for the closed string modes. In
particular, it is independent of the choice of closed string
vacuum. This is consistent with the fact that the boundary state
can be seen as the external source of all closed string fields,
which is clearly independent of the vacuum choice. Therefore, we
can expand $|B\ket$ in any basis.  For example, for later use we
now
 expand $\bra B|$   in terms of the basis $U^p_\om,V^p_\om$.

The mini-superspace relation \UVdef\ is modified in the exact
treatment of CFT \eqn\UVexact{\eqalign{
|U^p_\om\ket\ket&=|L^p_\om\ket\ket +{\cal
R}(p,\om)|R^p_\om\ket\ket \cr |V^p_\om\ket\ket&=|R^p_\om\ket\ket
+{\cal R}^*(p,\om)|L^p_\om\ket\ket }} where ${\cal R}(p,\om)$ is
the exact reflection coefficient \eqn\Rexact{ {\cal
R}(p,\om)={\cal R}_0(p,\om){\Ga(1+ipb^2)\o\Ga(1-ipb^2)} } From
\UVexact\ we can easily find the disk one-point function of
$U^p_\om,V^p_\om$ \eqn\UVone{\eqalign{ \bra
B|U^p_\om\ket&=\Psi^L(p,\om)+{\cal R}(p,\om) \Psi^L(-p,\om) \cr
\bra B|V^p_\om\ket&=\Psi^L(-p,\om)+{\cal R}^*(p,\om) \Psi^L(p,\om)
}}  Therefore in terms of the   basis \UVexact\ the boundary state
can be expressed as \eqn\BinUV{\eqalign{ \bra
B|=\int_{-\infty}^\infty{d\om\o2\pi}\int_0^\infty{dp\o2\pi}
{1\o1-|{\cal R}(p,\om)|^2}\Big[&\bra\bra
U^p_\om|\Big(\Psi^L(p,\om)-{\cal R}^*(p,\om)
\Psi^L(-p,\om)\Big)\cr &\bra\bra V^p_\om|\Big(\Psi^L(-p,\om)-{\cal
R}(p,\om) \Psi^L(p,\om)\Big)\Big] }}

\subsec{Annulus Amplitude and Radiation} Let us consider the
annulus amplitude between two D0-branes outside the horizon. After
coupling the black hole system with $D$ free bosons in such a way
that \eqn\Dcrit{ D+2+6b^2=26, } the imaginary part of the annulus
amplitude, calculated in the open string channel as before, turns
out to be
 essentially the same as the
hairpin brane case: \eqn\ImZabs{\eqalign{ {\cal Z}&=\bra B|{1\o
L_0+\til{L}_0+i\ep} |B\ket \cr &=\int_0^\infty{dt_o\o
t_o}t_o^{-{p\o2}}\eta(q_o)^{-D}\int dsdE q_o^{{E^2\o4k}+
{b^2s^2\o4}} {\sin{\pi E\o kb^2}\o\sin{\pi E\o k}(\cosh(\pi
s)+\cos{\pi E\o kb^2})} \cr \longrightarrow{\rm Im}{\cal
Z}&=k\int_0^\infty{dt_o\o t_o}t_o^{-{p\o2}}\eta(q_o)^{-D} \int
ds\sum_{n\in{\Bbb Z}}q_o^{{kn^2\o4}+ {b^2s^2\o4}} {(-1)^n\sin{\pi
n\o b^2}\o\cosh(\pi s)+\cos{\pi n\o b^2}} }} In particular this
imaginary part is UV finite   since \eqn\massinf{
\til{m}_\infty^2={k\o4}-{D\o24}=\qu\lf({1\o b}-b\ri)^2 \geq0. }

However, in computing the closed string radiation  we  have to be
careful about the vacuum choice which exists here due to the
presence of the horizon. This is most easily seen in the closed
string channel: when choosing a particular vacuum, the vacuum
independent source term $|B\ket$ is decomposed into positive and
negative frequency parts \eqn\Bpm{
|B\ket=|B^{(+)}\ket+|B^{(-)}\ket. } With the Feynman prescription
of the closed string propagator $(L_0+\til{L}_0+i\ep)^{-1}$, each
component $|B^{(\pm)}\ket$ contributes to the amplitude with a
definite sign. To see the effect of vacuum choice simply, let us
consider the particle limit of the amplitude \ImZabs, evaluated in
the closed string channel \eqn\Zsim{ {\cal Z}\sim \int
d^2x\,d^2y\,\bra B|x\ket G(x,y)\bra y|B\ket } Here we approximate
the boundary state by the mini-superspace wavefunction \eqn\Bmini{
\bra B|x\ket=\int{dpd\om\o(2\pi)^2}\Psi^L(p,\om)L^p_\om(x) } We
also set $D=0,k={9\o4}$ for simplicity. Namely, we consider the
purely two dimensional  black hole.

As discussed in \DijkgraafBA, the Green function $G(x,y)$ depends
on the vacuum. For example, in the Hartle-Hawking vacuum $G(x,y)$
reads (for $t_x>t_y$) \eqn\GHH{ G_H(x,y)=\int_{-\infty}^\infty
dp\rho(p)\Big[(1+N_\om)U_p(x)U_p(y)^*+N_\om U^*_p(x)U_p(y)\Big] }
where \eqn\Updef{ U_p=U^p_{\om=-3p>0},\quad U_p=V^p_{\om=3p>0} }
and $N_\om$ is the Boltzmann distribution function \eqn\Nomdef{
N_\om={1\o e^{2\pi\om}-1} } The measure $\rho(p)$ is given by
\eqn\rhop{ \rho(p)={\pi\o6}|\tanh2\pi p| } In this Hartle-Hawking
vacuum the closed strings are in thermal equilibrium with the
black hole, with temperature $T_H$. As the temperature of the
hairpin brane is lower than that, the system is not at equilibrium
and the radiation as seen by an asymptotic observer is not
thermal, as can be easily seen from the explicit form of the one
point functions.

 On the other hand, in the
Schwarzschild vacuum there is no asymptotic closed string
radiation and $G(x,y)$ has no thermal factor $N_\om$ \eqn\Gs{
G_S(x,y)=\int_{-\infty}^\infty dp \rho(p)U_p(x)U_p(y)^* } hence
the radiation seen by this observer comes entirely from the
D-particle and it is thermal (in the open string variables, as
explained in the previous section).

Therefore  we find that the radiation as seen by the Schwarzschild
observer is precisely what we obtain from the calculation in the
open string channel, and is very similar to the case of the
hairpin brane in flat space. Apparently the calculation in the
open string channel involves implicitly a vacuum choice for the
closed string sector. It is interesting to investigate whether the
open string calculation can yield more general such vacua.

\newsec{Discussion and Outlook}

There is no evidence of black hole formation in the high energy
scattering of closed string tachyons in the singlet sector of
$c=1$ matrix model \refs{\MooreZV,\MartinecQT,\KarczmarekBW}.
Therefore, it is widely believed that the 2d black hole is related
to the non-singlet sector of matrix model. In \KazakovPM, based on
the FZZ duality between the cigar and the sine-Liouville, it is
argued that the Euclidean 2d black hole is described by a matrix
model with Wilson loops added to the action. However, a direct
Lorentzian description of 2d black hole is still lacking. Our
result of D0-branes in Lorentzian black hole may shed light on
this problem. From the general philosophy of open-closed duality,
the Lorentzian black hole should be dual to the theory on a large
number of D0-branes on the black hole background. Our result
suggests that the Lorentzian black hole is dual to a system of
D0-branes and ghost D0-branes, {\it i.e.} $U(n|m)$ supermatrix
quantum mechanics.\foot{
Note that in \refs{\SenYV,\MartinecQT}
it is proposed that the black hole state in the $c=1$ matrix model
is represented by a large number of soft particle-hole pairs.
} In particular, the open string stretched
between D0-branes outside the horizon and D0-branes behind the
horizon is fermionic. Note that the appearance of fermionic open
string in bosonic string theory is not new. Indeed it is known that
the open string
between FZZT-brane and ZZ-brane in bosonic minimal string theory
is fermionic \KutasovFG. This is also consistent with the proposal
that the effect of Wilson loops in the KKK matrix model is
reproduced by introducing non-singlet fermions into a Lorentzian
matrix model \refs{\MaldacenaHI,\GaiottoGD}. We leave the study of
supermatrix model and its relation to 2d black hole as an
interesting future problem.

\vskip 5mm \noindent {\bf Acknowledgments:} KO would like to thank
Yuji Sugawara for correspondence, MR thanks M. Berkooz and R.
Myers for interesting conversations. MR is supported by an NSERC
discovery grant. Research at the perimeter institute is supported
in part by funds from NSERC of Canada and MDET of Ontario.

\listrefs
\bye